# *The Evolution of the Computerized Database*

Nancy Hartline Bercich

Advisor: Dr. Don Goelman

Completed: Spring 2003

Department of Computing Sciences

Villanova University

Villanova, Pennsylvania

U.S.A.

# Database History

## *What is a database?*

Elsmari and Navathe [ELS00] define a database as a collection of related data. By this definition, a database can be anything from a homemaker's metal recipe file to a sophisticated data warehouse. Of course, today when we think of databases we seldom think of a simple box of cards. We invariably think of computerized data and their DBMS (database management systems).

## *Non Computerized Databases*

My first experience with a serious database was in 1984. Asplundh was installing Cullinet's IDMS/R™ (Integrated Relational Data Management System) manufacturing system in the company's three manufacturing facilities. When first presented with the idea of a computerized database keeping track of inventory and the manufacturing process, the first plant manager's response was, " Why do we need that? We already have a system. We already have a method for keeping track of our inventory and manufacturing process and we like our method. Why change it?" He retrieved, from a metal filing cabinet, several boxes containing thousands of index cards. Each index card recorded data on a specific part, truck, manufacturing process or supplier. He already had a database, a good database that quite possibly predated the modern computer. In the end our task we easy. We simply mechanized his existing database.

## *Flat File Storage*

Before the advent of modern database technology computerized data was primarily stored in flat files of varying formats, ISAM (Indexed Sequential Access Method) and VSAM (Virtual Storage Access Method) being two of the more common file formats of the time. Unfortunately, there are several disadvantages inherent in the use of flat files. First, flat files do not link data in one file to data in another file in a mechanized way. Linking the data in one file to the data in another file is the responsibility of the software developer. This complex task involves defining and opening each file; accessing data in the first file, coding data access paths into the program and finally accessing the second file. As these files are fixed-format files, modifying the structure of any one flat file is a rather lengthy process. Modification of a file structure first requires conversion of the data in that file. Modification next requires modifying the flat file definition in every program accessing the file, and possibly modifying the file access paths in each of these programs as well. Failing to modify a single program could result in incorrect reporting, premature program termination or possibly even data corruption. In order to overcome the errors resulting from the work born of this limitation, the same data is often stored in multiple flat files. This creates a second problem, data redundancy and the data inconsistencies, which inevitably come about because of this practice. The search for a resolution to these data redundancy, inconsistency and access problems led to the development of the various data models used today. This is not to say there are not instances where flat files provide an optimal storage solution. Creating and maintaining flat files in a stable production environment is easier than creating and maintaining a DBMS file system. Creating and maintaining flat files under these conditions is also inexpensive when compared to the cost of creating and maintaining a DBMS file system. Commercial DBMSs are expensive to purchase and maintain. This expense comes from both licensing and maintenance fees as well as personnel costs. As DBMSs become more complex, their licensing and maintenance fees grow higher still. Well-trained and experienced DBMS software professionals command higher salaries than the less specialized software professionals needed to properly develop and manage a flat file system. DBMS personnel are often not available due to the steep learning curve on some DBMSs. This requires a company to put their existing software personnel through extensive training only to have them move on once trained.

## *Computerized DBMSs*

Today DBMSs are categorized by their underlying implementation data model. An implementation data model describes the structure of a database. It describes the manner in which a database stores and links data together. There are several distinct database models adhered to by the more common DBMSs commercially available today. These include the network, hierarchical, relational, object and object-relational models. The most widely used of the modern databases currently adhere to either the relational or object-relational data model. Today we find database systems running on DBMSs that adhere to either the network or hierarchical database models in legacy systems. The oldest of these models are the network and hierarchical models, the newest the object and object-relational models. There are of course other, less common, more specialized, models around as well. [ELS00] Carefully reviewing each of the more common data models can provide a picture of the evolution of the DBMS to date. Further review may provide a look at the future evolution of the DBMS.

## Network and Hierarchical Databases

The first attempt at organizing data into databases resulted in the development of the network data model. The CODASYL (Conference on Data Systems Languages) Data Base Task group first proposed the network data model in 1971. DBMSs, such as IDMS/R™, based on the network data model, organize data into records with each record containing a group of related data items or attributes. The term record type refers to the structure of a record. Set types connect records, where a set type represents a 1:N relationship between two record types. Each set type definition consists of three elements, a set type name, an owner record type and a member record type. In a network database records are accessed one record at a time using a DML (data manipulation language) embedded in a host programming language. The most commonly used host languages being COBOL (Common Business Oriented Language) and PL/1 (Programming Language 1). The database network is traversed one record at a time through a record type or from one record of one record type to another record of a second record type through a set type. [UNIX]

The hierarchical data model, developed around the same time as the network data model, is similar, in many respects, to the network data model. As in the network data model, the hierarchical data model organizes data into records with each record containing a group of related data items, referred to as fields. Record type again refers to the structure of a record. The hierarchical data model differs from the network data model in the manner in which it links record types. Hierarchical DBMSs, such as IBM's IMS™ (Information Management System), organize data in a tree structure, beginning with a single record type at the top of the tree structure. There is a hierarchy of parent and child record types, with each parent child relationship being a 1:N relationship. Hierarchical databases access data using an HDML (hierarchical data manipulation language) a single record at a time. The HDML accesses records by moving from the current record to the next record in a path or even several records down the path depending on the HDML command used. [UNIX]

DBMSs based on the network and hierarchical data model provided software professionals, working with large quantities of data, a degree of control they did not have when working with flat files. DBAs (Database Administrators) could do more than just maintain data. DBAs could manage data. Both models saved the links between records of differing record types along with the data contained in those records. This freed the software developer from having to recreate the data links with every new program, minimizing data traversal mistakes. In addition, this ability, to move from one record type to another, minimized the need to store commonly used data in multiple locations, reducing data redundancy and hence data corruption.

However, DBMSs based on the network and hierarchical data model were not flawless. Accessing data in either a network or hierarchical database is rigid, time consuming and difficult. As hierarchical databases do not allow links between any of two child record types, a user has to start at the top of the data hierarchy and work down through the database structure. See Appendix A, Figure 1. The user needs to know and have access to the entire database structure to use any part of the database. Although searching for data in a network database is slightly less rigid, as network databases allow child record types to be linked to each other, accessing data deep in the network is still time consuming and difficult. See Appendix A, Figure 2. Codd [COD70] points to several more flaws in his 1970 paper, "A Relational Model of Data for Large Shared Data Banks." DBMSs based on the network and hierarchical data models lack independence of application programs from growth in data types and changes in data representation. A considerable amount of coding could still result from a minor modification to a record type definition. In addition, three types of data dependencies exist in DBMSs adhering to the network and hierarchical data models: ordering dependence, indexing dependence, and access path dependence. Network and hierarchical databases do not make a clear distinction between the order of presentation and stored order. Thus, a change in stored order of data often requires a change in the programs accessing that data. With some commercially available hierarchical databases, such as IDS (Integrated Data Store), developers must modify application programs and terminal activities as indices come and go. Others, such as IMS™ and IDMS/R™ do not have this limitation. However, if the developer or DBA changes the structure of the trees or networks in these systems, previously developed application programs tend to become logically impaired.

Network and hierarchical DBMSs were popular from the late 1960s through the 1970s and even into the early 1980s. IBM's IMS™ DBMS was the most popular of the hierarchical databases and Cullinet's IDMS/R™ the most popular of the network databases. Although relational databases are more commonly used in systems written today, databases based on both the network and hierarchical model are still in use today, mostly in legacy systems running on large mainframe computers.

## Relational Databases

E.F. Codd [COD70] first defined the relational database model in a 1970 article for ACM (Association for Computing Machinery) entitled "A Relational Model of Data for Large Shared Data Banks." In this article, Codd detailed what he felt were deficiencies in the existing DBMSs adhering to the network and hierarchical data model. Codd's relational model uses, as its basis, the mathematical relation; set theory and first order predicate logic. Codd further refined the relational model in two more papers written in 1985. In these two papers, he spelled out twelve rules to which a DBMS must adhere in order to be relational. These twelve rules invariably address the deficiencies Codd found in the industries earlier attempts at developing commercial DBMSs. Codd's rules are the information rule, the guaranteed access rule, the rule requiring systematic treatment of null values, the rule requiring the existence of a dynamic on-line catalog based on the relational model, the comprehensive data sub-language rule, the view updating rule, the high-level insert, update, and delete rule, the physical data independence rule, the logical data independence rule, the integrity independence rule, the distribution independence rule and the non-subversion rule. [COD85][COD85a]

It is interesting to note that not one of the existing DBMSs today, that are referred to as RDBMSs (relational database management systems), conform to all twelve of Codd's rules. Still, we can best describe the advantages of these RDBMSs over the network and hierarchical DBMSs by examining a few of these twelve rules. First, by following the guaranteed access rule, an RDBMS allows easy access of all data. See Appendix A, Figure 3. Gone are the path requirements of the network and hierarchical databases. The relational model provides a means of describing data with its natural structure only. Network and hierarchical databases require additional structure for machine representation. Next, by following the physical and logical data independence rule RDBMSs remove the necessity of modifying application programs whenever modifying the structure of the database. Finally, by following the integrity independence rule RDBMSs minimize data corruption.

Even with these improvements, RDBMSs are far from perfect. Elsmari and Navathe note that the now traditional RDBMS data model and RDBMS database systems have certain shortcomings when designing and implementing more complex database applications. They note that newer applications have requirements for transactions of longer duration and new data types for storing images, large text items and complex objects. [ELS00].

Databases adhering to the relational model have been the dominant databases since the middle of the 1980's with Oracle v6 and v7 the best selling of the relational databases.

## Object-Oriented and Object-Relational Databases

OODBMSs (object oriented database management systems) began appearing in the early 1990s, largely to meet the needs of the more complex applications described above. The first attempt to define an OODBMS standard occurred in 1993 when Malcolm Atkinson, François Bancilhon, David DeWitt, Klaus Dittrich, David Maier and Stanley Zdonik [ATK93] published their paper, "The Object-Oriented Database System Manifesto." Since that time, the ODMG (Object Data Management Group) has taken over this task. The ODMG 3.0 object data standard defines specifications for storing objects in OODBMSs, as well as for converting and storing objects in Object-Relational databases. The ODMG based their ODMG 3.0 object data standard on OMG's (Object Management Group) object model. [CAT99] The concepts behind the OMG object model are the same concepts behind all object-oriented technologies, encapsulation, polymorphism and inheritance. An OODBMS stores two types of data, objects and values. An object typically has two components, state and behavior, where the state is represented by values or previously defined objects and the behavior by operations on those values and objects. An OODBMS supports encapsulation, the association of data with the code that manipulates that data, by storing both components together in the database. It supports polymorphism, a mechanism associating one interface with multiple code implementations, by the manner in which it allows operation definition. Finally, it supports inheritance, a mechanism allowing one object to be the specification of another object, by allowing an object's state to consist of both values and other, previously defined objects. [ONE99] Objects in object-oriented programming languages exist only for the duration of the program. An OODBMS allows these objects to persist beyond the program's execution time.

OODBMSs have several advantages over their predecessors. They allow for the persistent storage of complex data types. This allows developers using object-oriented programming languages, such as Java™ and C++, to store and retrieve instances of their objects in databases, without disassembling them into their relational parts. This reduces the risk of programmer error when reconstructing a stored object. OODBMSs also, by allowing for the storage of an object's behavior, reduce the time needed to modify front-end programs when modifying an object's behavior. Because all programmers use the same operations to access an object's state, unless a developer is modifying an operation's interface, he only needs to modify a single stored operation. See Appendix A, Figure 4. He even avoids the necessity of program recompilation. Yet, even OODBMSs are not perfect. The ODMG 3.0 standard states that no existing OODBMS supports complete object encapsulation. [CAT99] This limits the advantages gained from storing both an object's state and behavior.

Object-Relational databases are, as their name implies, a hybrid of object and relational databases. These databases grew from the efforts of the RDBMS vendors to address the deficiencies in the relational model. There is no standard definition of an object-relational database. There is only the ODMG standard for ODMs (object-to-database mappings) that convert and store objects in relational representations. Object relational databases do not even follow this standard, as object-relational databases do not attempt to map objects to relations. Object-relational databases allow for the storage of complex objects. Oracle 8i and Oracle 9i are two examples of object-relational databases. Oracle 8i and Oracle 9i both allow for the storage of complex objects. Oracle 9i allows for object inheritance. Neither allows an object's behavior to be stored along with its state. [LOR99] [FRE02]

OODBMSs generated significant interest, in both the research and business communities, in the early to the middle of the 1990s. However, in recent years interest, from both communities, has waned. Object-relational databases now hold a significant portion of database market share, largely due to users upgrading their previously relational database to the vendor's newer object-relation database. The most dominant of these object-relational databases are Oracle 8i and Oracle 9i.

# Industry Trends

## *Introduction*

It's human nature to always look for an easier and better way of doing things. Computer Scientists are no different from anyone else. New technologies, in the field of Computer Science, will always be on the horizon. Still, the time it takes for a new technology to move from a new force to a dominant force is not a matter of weeks. Often, it is a matter of years or even decades. E.F. Codd published his first paper on RDBMSs in 1970. He refined his definition of RDBMSs in 1985. ORACLE®, according to their own web site, released version 2 of Oracle in 1980 and version 5 in 1986. ORACLE® did not begin to establish itself the dominant, commercially available, DBMS until the release of Oracle version 6 in 1988. [ORA]

It takes time for a new technology to establish itself for a myriad of reasons. Whenever a new technology appears there will always be those who embrace the new technology and those who reject the new technology. There are advantages and disadvantages to both approaches to dealing with new technology. Looking at this issue from a business viewpoint, timing is critical. The adoption of a new technology requires a significant investment in both software and training. Early adopters run the risk of finding themselves with a large investment in an obsolete or unsupported technology. Not all, new technologies are eventually adopted by the industry. Even if a new technology is eventually, adopted, not all vendors' products, adhering to the new technology, will survive. Developing and marketing a software product requires a significant investment in time and money and not all vendors can stay the course. Late adopters risk facing another problem. Late adopters may find themselves scrambling to catch up when a new technology provides their competitor with a significant business advantage. Businesses must carefully weigh the pros and cons of adopting a new technology. Businesses must wait for the right moment and the right product. For software professionals, the decision to learn a new technology is a far easier decision to make. With many, though not all, new technologies, the only investment the software professional needs to make is an investment in time. It takes time to learn a new technology. The only disadvantage is that the time a professional spends learning one new technology he cannot spend learning another.

## *What happened with RDBMSs*

In the early to middle of the 1980s when RDBMSs first became commercially available, many in the industry failed to see the advantages of switching to the newer technology. For others, the risks were just too great. For a brick and mortar business, the decision on whether or not to embrace the new technology was a difficult decision to make. Adopting the new technology required a significant financial investment, an investment in software, software conversion, employee training and possibly even hardware. The older, network and hierarchical databases worked. The newer RDBMSs, although eliminating many of the undesirable traits of the network and hierarchical databases, initially ran slowly. The decision as to which RDBMS to implement was equally as difficult. Many perceived IBM®, the dominant force in the computer industry as strictly a hardware company. Buying software from IBM® seemed risky. ORACLE® Corporation was an unknown entity that many worried would fold. Making an early decision to go with an RDBMS that failed to remain marketable would cause the company to incur significant financial losses. For a business whose products were software professionals, such as a consulting firm the decision was less difficult. For these businesses, the need to stay technologically current far outweighed the risk involved in investigating an unproved technology. Businesses such as these incurred a smaller debt when investigating and adopting the newer technology. The licensing fees they were required to pay were significantly smaller than those paid by brick and mortar businesses. They could retrain their employees and tout the benefits of the new technology. At the same time, they could keep their own systems on older, more stable technologies, while waiting for the market to adopt the new technology and a market leader to emerge. An initial decision to go with a vendor whose product failed to gain market share was not a disaster for one of these firms. The firm that employed me initially bet on DB2™. IBM® was an established company with an established customer base. It would appear with ORACLE® eventually became the dominant RDBMS vendor, that this was a bad decision. It was not. Those of us initially trained in DB2™, were able to easily relate the theories we learned to the study of ORACLE®, thereby reducing our learning curve when it became apparent that ORACLE® was the RDBMS to know. Consulting firms who trained their network and hierarchical database professionals in relational technology, early on, found their services in great demand when the newer relational technology became dominant.

### *What is happening now*

Cullinet® marketed IDMS/R™ as relational although it was definitely not. In 1980, everything that was anything was relational. By 1985, database professionals who were not learning relational technologies were worried. Today, everything that is anything is object oriented. . Computer professionals are learning Java™, UML (unified modeling language) and OO (object oriented) design techniques. Database professionals who, for the most part, are still working with RDBMSs, are concerned that they are becoming dinosaurs. Those few with OODBMS knowledge are concerned when they are not working with OODBMSs. The industry, as a whole, is embracing the OO paradigm. It is not yet clear; to what extent the database industry will follow suit.

### What about Object-Relational

It would appear, with the popularity of Oracle 8i and 9i that object relational databases are the future of database technology. However, we cannot assume that's the case. As ORACLE® Corporation releases newer object-relational versions of its popular RDBMS, it discontinues support for it's older strictly relational versions. There is no way to tell how many users of these newer object-relational databases are actually using the object functionality. Many may merely be using an object-relational database because a vendor has dropped support for the older RDBMS to which they have tied themselves. ORACLE® has begun selling a scaled down version of their Oracle 9i database. It does not contain the object functionality. Many existing Oracle 8i customers are upgrading to that version of Oracle 9i.

### What about OODBMSs

Although object technologies such as SMALLTALK have been around since the early 1970s it has only been in recent years that object design tools such as UML diagramming and programming languages such as Java™ and C++ have become universally popular. The first attempt at defining an OODBMS standard did not occur until 1993. [ATK93] Versant, one of the longer established OODBMS vendors only established itself in 1988. [VERS] In the middle of the 1990s, it appeared as if OODBMSs were in position to take market share away from RDBMSs. Companies began purchasing OODBMSs. Industry professionals began training themselves in the OODBMS technologies. Then, the popularity of OODBMSs, as indicated by new market share, began to decline.

### *What next*

It's not clear in which direction the DBMS industry is heading. Many thousands of businesses have tied themselves to one particular DBMS. Although the relational model has clearly shown its superiority to the older network and hierarchical database models, many of the businesses that first embraced database technology are still running legacy systems in IMS™ and IDMS/R™. Their reasons are purely financial. Thousands more businesses initially tied themselves to the newer RDBMSs than to the older, network and hierarchical, DBMSs. The OO paradigm, in non-database technology, is becoming the industry standard. Businesses, as they now embrace this paradigm in other areas of software development, have a decision to make. They need to decide whether to stay with their current RDBMS, which even as it becomes object-relational, still lacks complete support for storage of the transient complex objects they are developing or spend the money to move from an RDBMS to an OODBMS, which would fully support their objects. Despite their obvious advantages, when dealing with complex objects, OODBMSs have not otherwise shown clear superiority over RDBMSs. An OODBMS market leader has yet to emerge. For a brick and mortar business, investing in an OODBMS is still risky. The object data model may yet show superiority over the relational model. If it does, brick and mortar businesses will begin to embrace the new technology. A market leader will emerge. Software professionals who have taken the time to learn OODBMS skills will be in high demand. If the object data model fails to show clear superiority over the hybrid object-relational model, those same professionals will have lost nothing but time. For software professionals, understanding the new object data model is smart business.

# Appendix A: Sample Diagrams

## *Hierarchical Occurrence Tree*

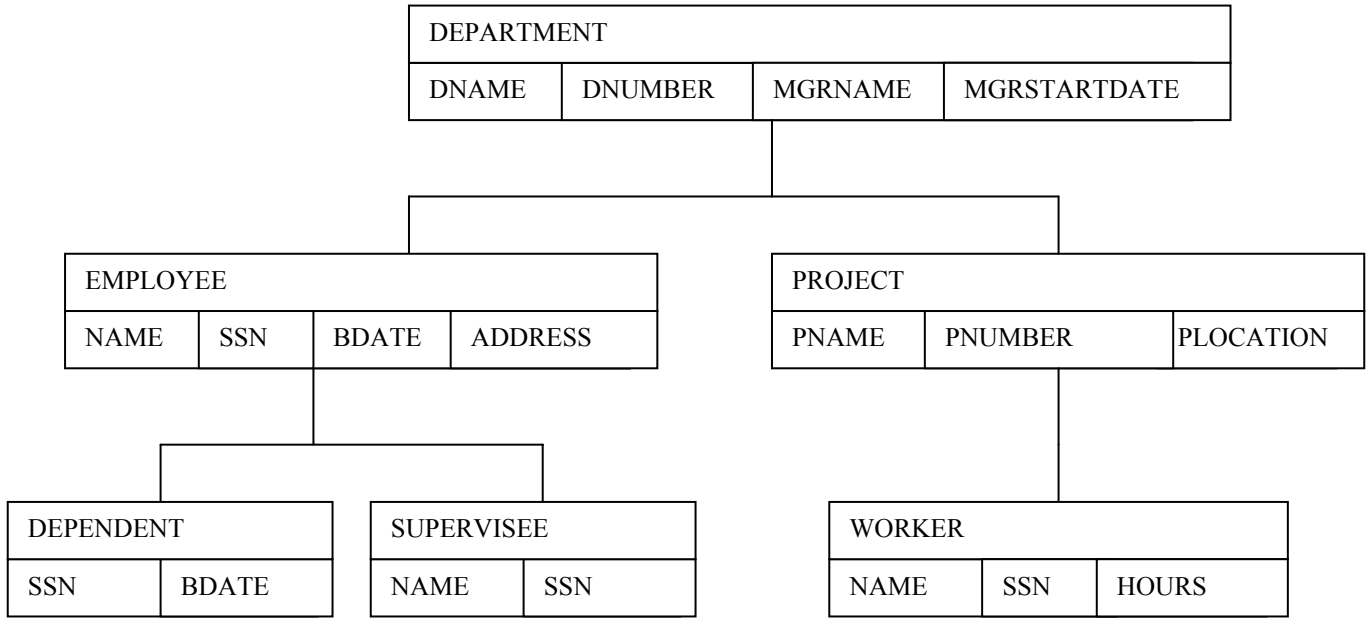

**Figure 1:** The figure shown above, taken and modified, from Elmasri and Navathe p. 945 shows a hierarchical occurrence tree representing the record layout of a hierarchical COMPANY database. Note that there exists a single entry point at the top of the data hierarchy. Note also how the hierarchical design forces record access down through the database along the edges shown in the diagram. [ELM00]

## *A Simplified Network Schema Diagram*

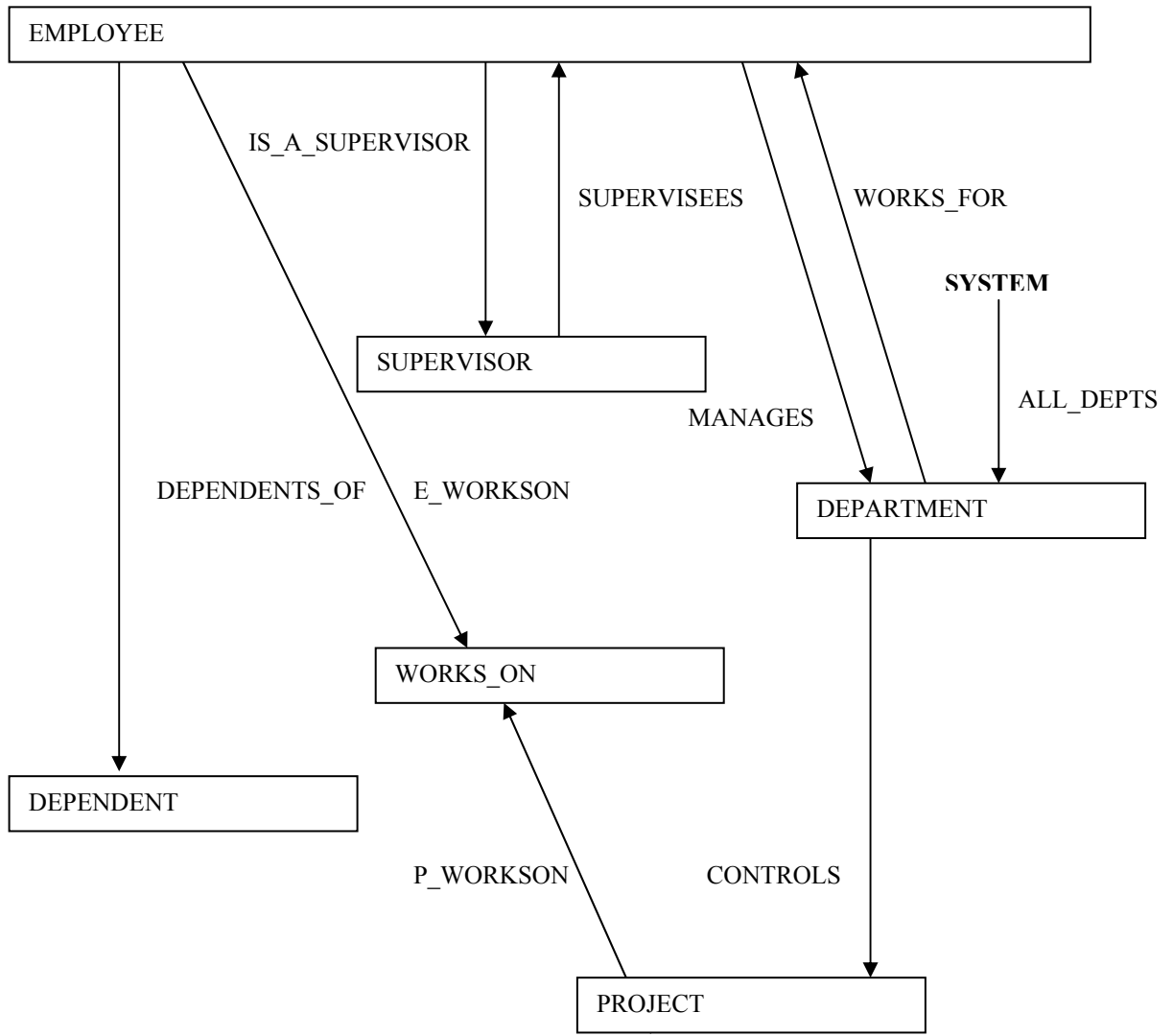

**Figure 2:** The figure shown above, taken, and modified, from Elmasri and Navathe p.931 shows a simplified network schema diagram for a COMPANY database. This simplified network schema diagram shows only the records that would appear in the database as well as the navigation through the database. Notice the single entry point into the database denoted by the label SYSTEM. Notice also how traversal of the database, represented diagrammatically with arrows, is limited. [ELS00]

## *ER Diagram*

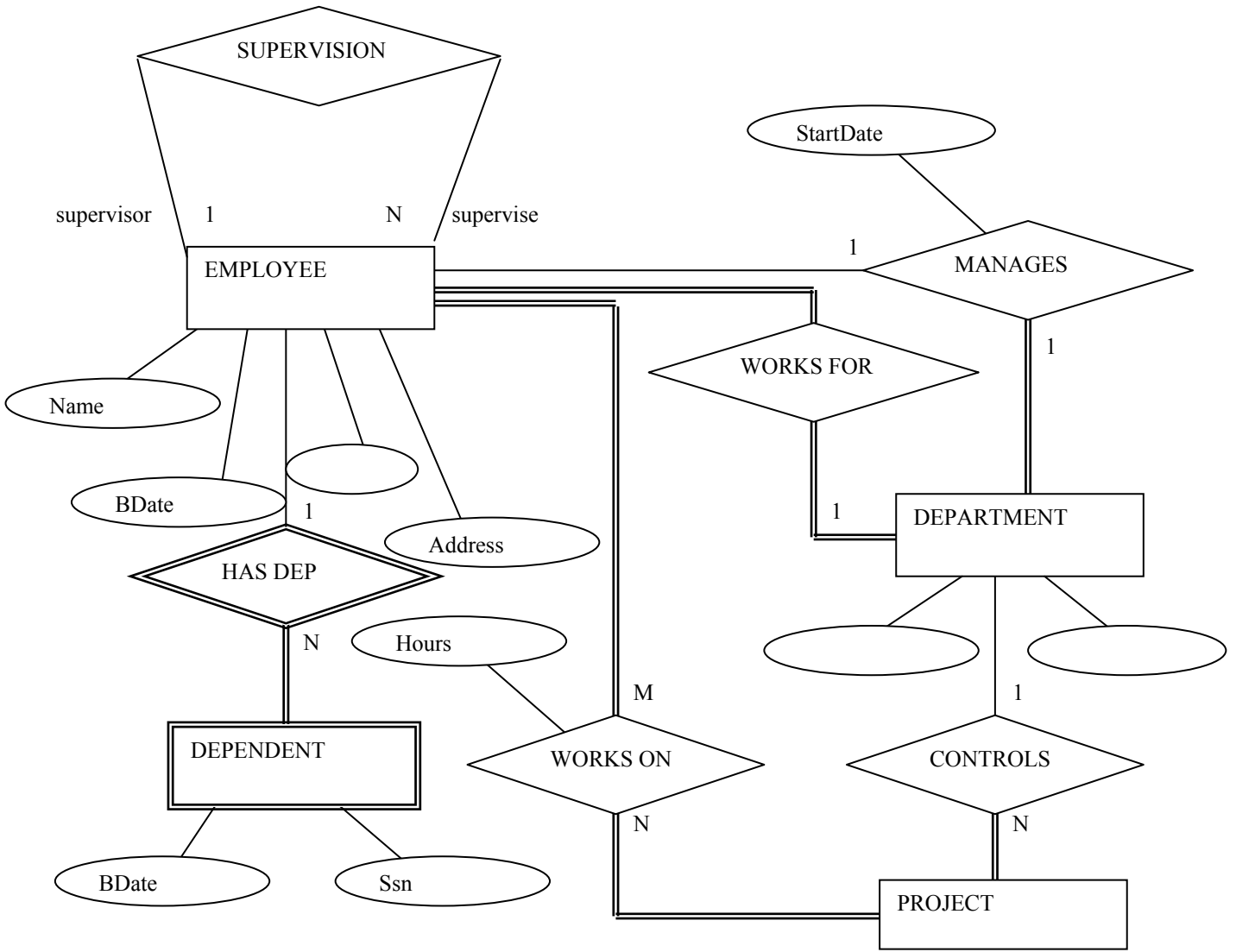

**Figure 3**: The figure shown above, taken, and modified, from Elmasri and Navathe p.46 shows an ER schema diagram representing an RDBMS COMPANY database. Notice the absence of the database entry point and the navigational arrows present in the network schema diagram. Navigation, within an RDBMS, can begin at any table and move in any direction to any associated table allowing easy access of all data. [ELS00]

UML Conceptual Schema

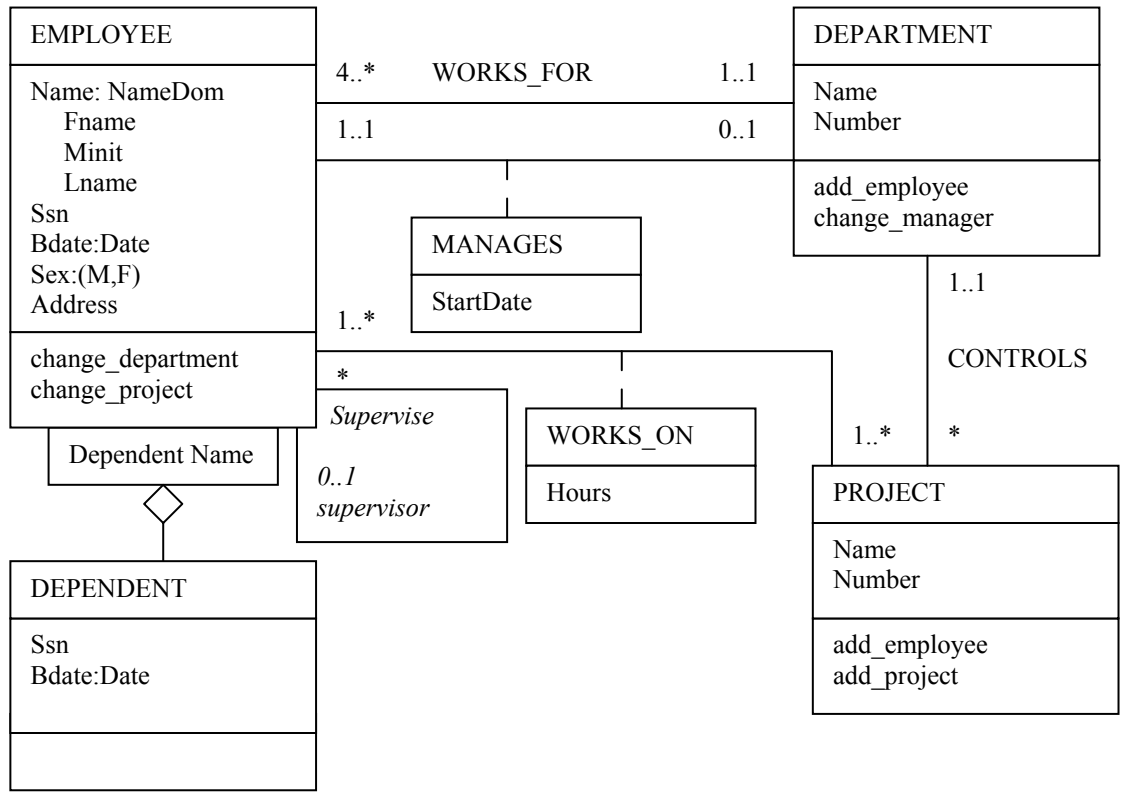

**Figure 4:** The figure shown above taken, and modified, from Elmasri and Navathe p.94 shows a UML conceptual schema for a COMPANY database. A UML conceptual schema diagrammatically represents the format of an OODBMS. Notice how the methods operating on the objects are associated with the objects. [ELS00]

# Bibliographic References